\documentclass[a4paper,fleqn,usenatbib]{mnras}

\usepackage{newtxtext,newtxmath}

\usepackage[T1]{fontenc}
\usepackage{ae,aecompl}

\usepackage{graphicx}	
\usepackage{amsmath}	
\usepackage{amssymb}	
\usepackage{ulem} 

\title[Modelling SED of super-Eddington quasars]{Modelling the spectral energy distribution of super-Eddington quasars}

\author[A. Kubota \& C. Done.]{
Aya Kubota$^{1,2}$\thanks{E-mail: aya@shibaura-it.ac.jp }
and Chris Done$^{2}$
\\
$^{1}$Department of Electronic Information Systems, Shibaura Institute of Technology, 307 Fukasaku, Minuma-ku,   Saitama-shi, Saitama 337-8570, Japan\\
$^{2}$Department of Physics, University of Durham, South Road, Durham, DH1 3LE,UK\\
}

\date{Accepted XXX. Received YYY; in original form ZZZ}

\pubyear{2019}

\begin{document}
\label{firstpage}
\pagerange{\pageref{firstpage}--\pageref{lastpage}}
\maketitle

\begin{abstract}

We develop a broadband spectral model, {\sc agnslim}, to describe super-Eddington black
hole accretion disc spectra. This is based on the slim disc
emissivity, where radial advection keeps the surface luminosity at the
local Eddington limit, resulting in $L(r)\propto r^{-2}$ rather than
the $r^{-3}$ expected from the Novikov-Thorne (standard,
sub-Eddington) disc emissivity. Wind losses should also be important
but these are expected to produce a similar radiative emissivity.  We
assume that the flow is radially stratified, with an outer standard
disc, an inner hot Comptonising region and an intermediate warm
Comptonising region to produce the soft X-ray excess. This gives the
model enough flexibility to fit the observed data, but with the
additional requirement of energy conservation to give physical
constraints. We use this to fit the broadband spectrum of one of the
most extreme Active Galactic Nuclei, the Narrow Line Seyfert 1
RX~J$0439.6-5311$, which has a black hole mass of $(6\sim9)\times
10^6M_\odot$ as derived from the H$\beta$ line width. This cannot be
fit with the standard disc emissivity at this mass, 
as even zero spin models overproduce the observed luminosity. Instead, we show that the 
spectrum is well reproduced by the
slim disc model, giving mass accretion rates around $(5\sim 10)\times$Eddington limit.
There is no constraint on black hole spin as the efficiency is reduced by advection.
Such extreme accretion rates should be characteristic of
the first Quasars, and we demonstrate this by 
fitting to the spectrum of a recently discovered super-Eddington
Quasar, PSO~J$006+39$, at $z=6.6$.

\end{abstract}

\begin{keywords}
black hole physics -- galaxies: Seyfert -- accretion, accretion discs
\end{keywords}



\section{Introduction}

Active galactic Nuclei (AGN) are observed to shine at super-Eddington
accretion rates. This is not immediately apparent from using a constant
bolometric correction factor for the optical emission 
(e.g., \citealt{steinhardt2010}).
However, the accretion disc equations
do not predict a constant bolometric correction factor as the maximum
disc temperature increases with increasing mass accretion rate. Instead, these
give $L_{\rm opt}\propto (M\dot{M})^{2/3}\propto (M L_{\rm bol})^{2/3}$ \citep{collin2004, davis2011}.
\citet{collin2004} used this to show that some Narrow
Line Seyfert 1 (NLS1) AGN had optical spectra which implied that their
mass accretion rate was up to 10 times Eddington limit.  \citet{schulze2017}
derive the Eddington ratio from a large sample of AGN using both techniques, and show that 
the disc equations transform the distribution from having
a clear maximum at Eddington limit to extending significantly above Eddington limit.
This observational support for exceeding the Eddington
limit is important as the discovery of quasars with black hole masses
of $\sim 10^9 M_\odot$ at $z>7$ (e.g. \citealt{mortlock2011,banados2018})
puts stringent constraints on the mass of the initial black hole
seed. There is not enough time for such massive black
holes to grow from even $100 M_\odot$ pop III stellar remnants by
Eddington limited accretion for any reasonable black hole spin
\citep{shapiro2005}.  Either the seed black holes are more massive, though 
how these could form is not yet clear (e.g. \citealt{boekholt2018,wise2019})
or super-Eddington rates are required. 

However, merely having a super-Eddington luminosity does not
necessarily mean that the black hole can accrete all this material.
The accretion flow could respond by ejecting a powerful wind, as
originally suggested by \cite{ss73}. Such winds could be the origin
of AGN feedback, setting the $M_{\rm BH} -\sigma$ relation \citep{king2003},
but the wind losses reduce the mass accretion rate onto the central
black hole to around Eddington limit, so this does not solve the problem of
black hole growth in the early Universe. Nonetheless, there is an
alternative way for the accretion flow to exceed the Eddington
limit. The \cite{ss73} equations assume that the energy released
through viscosity is radiated locally, but this is not true close to
Eddington limit. The flow becomes very optically thick at high mass
accretion rates, and photons produced close to the mid plane do not
have time to diffuse vertically through the flow in order to be
radiated from the surface before the flow itself has moved radially
inwards. This optically thick advective cooling by photon trapping is necessarily
important in Shakura-Sunyaev discs around Eddington limit. It results in a
flow which is moderately geometrically thick, with $H/R\sim 1$ where $H$ is the scale height and $R$ is radius
\citep[slim discs]{abramowicz88}.  The radiation escaping from the
surface of the flow is limited to the local Eddington flux, but all
the additional super-Eddington radiation flux and super-Eddington mass
accretion rate is accreted, fueling maximum growth of the black hole.
Winds could co-exist with advective flows (e.g. \citealt{poutanen2007}),
but are not required. Numerical simulations do show both occurring in
super-Eddington flows but the ratio of power between these two
processes is not clear (e.g. \citealt{kitaki2018} and references therein).  However,
both advection and winds have rather similar effects on the emitted spectrum, as they both
lead to the emitted flux being limited to the local Eddington flux.
We thus consider advective cooling 
alone, and use the data to guide our thinking.

It is clear that AGN spectra in
general are more complex than expected from simple blackbody disc
models. This makes them different to the stellar mass black hole
binaries, where the spectra can often be well fit by an optically
thick disc component, though even the most disc dominated spectra show
a small tail of emission to higher energies \citep{rm06,dgk07}.  However, in AGN, there are no
spectra which are as disc dominated as those seen in the black hole
binaries.  The optical/near UV generally does look blue enough to be
the outer disc, especially when dust corrections are taken into
account (e.g. \citealt{davis2007,baron2016}),
but the soft X-rays
appear to be dominated by an optically thick, warm Comptonised component
of $kT_e\sim 0.2\sim1$~keV, while the higher energy X-ray emission requires
an additional much hotter, optically thin region (e.g. \citealt{elvis1994}).
While the origin of these components are not understood, they
are surely powered ultimately by gravitational energy released by the
accreting matter.  Since there is something that looks like an outer
disc in the optical/near UV then we can use this to measure the mass
accretion rate powering this emission. 
The disc equations assume that mass accretion rate is constant with
radius, so having measured its value in the outer disc sets the 
luminosity of the disc at all radii via the
Novikov-Thorne (NT) emissivity \citep{nt73}.

The observed spectral complexity can be parameterised by a radial
stratification of the flow. Thermalisation is more or less complete in
the outer disc, giving blackbody emission, but inwards of some radius
the flow heating concentrates towards the upper layers, making a
sandwich geometry which hardwires the soft X-ray excess spectral index
to that observed \citep{petrucci2018}.  Then at some smaller radius
the flow evaporates completely, making the hard X-ray coronal region
(the {\sc agnsed} model in {\sc xspec} by \citealt{kd18} (hereafter KD18), an upgrade
of the original {\sc optxagnf} model of \citealt{done2012}). This
approach effectively combines a physically based emissivity
prescription which sets the overall energetics, with enough 
free parameters to fit the data.

However, these models cannot fit to super-Eddington flows, as the
advection/wind losses from the inner disc reduce its emitted
luminosity compared to that required to power the outer disc. Again,
this is most clearly seen in the extreme Narrow Line Seyfert 1 AGN,
where the mass accretion rate required to fit the observed (very disc
like) optical/UV continuum predicts far too much luminosity at higher
energies for any reasonable black hole mass \citep{jin2016,done2016,jin2017}
The alternative is to fit with
slim disc models, but while super-Eddington sources are actually the
most disc like of all AGN (see e.g. \citealt{pounds1995,jin2012,done2012}), 
they still show the warm and hot
Comptonisation regions which make them difficult to fit with purely blackbody
disc emission models \citep{mineshige2000,puchnarewicz2001}.

Here we extend our previous three-radial zone approach to super
Eddington flows, to build the first tractable physical spectral model
which can fit the data. We simply replace the NT
emissivity (which is only appropriate for sub-Eddington flows) with an
emissivity which changes from $L(R)\propto R^{-3}$ to $R^{-2}$ when
the local disc flux exceeds the Eddington limit to effectively
describe the effects of either advection or winds (or both) in
suppressing the luminosity of the inner disc. We show that this can
fit the best dataset of a highly super-Eddington AGN,
RX~J$0439.6-5311$, where previous attempts with {\sc optxagnf}
failed \citep{jin2017}. Energy loss via advection and/or winds
decreases the luminosity from the inner disc relative to that expected
from the outer disc, enabling slim disc fits to the data with
reasonable black hole masses.

We also demonstrate the utility of the code on much more limited
datasets from high redshift quasars. We are able to fit the (rest
frame) optical/UV spectrum of PSO J006+39 at $z=6.6$ \citep{tang2019} 
and confirm that this is most likely a super-Eddington flow. We discuss the role of the flow geometry in making the unusually blue optical/UV spectrum seen from this quasar, 
but find that an inner funnel would also increase the apparent luminosity from geometric beaming. This reduces the intrinsic mass accretion rate to below Eddington limit, in conflict 
with the assumption that the inner flow forms a funnel due to super-Eddington mass accretion. More data for this quasar, especially including its X-ray emission, are
needed in order to understand the origin of its unusual properties.

\section{Overall emissivity of the disc model}
\label{sec:overall}

The Eddington limit is straightforward to define in stars, as they are
spherical. The radially outwards radiation force is balanced by the
radially inward gravity. The disc geometry is much less
straightforward. Sub-Eddington (thin) discs extend down to the
innermost stable circular orbit, $R_{\rm isco}$, have the radial
component of gravity balanced mainly by rotation, while the vertical
component $GM/R^2 (H/R)$ is balanced by total pressure (gas plus
radiation) setting the scale height, $H$, of the disc.  As
$L\to L_{\rm Edd}$ there are multiple other terms which become important
e.g. radial pressure terms become important, and $H\to R$ so rotation
is sub-Keplarian.  Radial advection of heat becomes important, so not
all the energy released is radiated locally, and the inner edge of the
flow is not necessarily set by $R_{\rm isco}$ \citep{abramowicz88, watarai2000, heinzeller2007,abolmasov2015}.
Nontheless, we can get use the thin disc equations to gain some
physical insight into the results of more detailed numerical studies
(see also \citealt{abramowicz2005} for an insightful review).

The overall emissivity for a sub-Eddington disc is given by NT 
as 
\[
F_{\rm NT}= \frac{3 G M \dot{M} f(r,a^\ast) }{8\pi R^3} 
=\frac{L}{4\pi R^2} \frac{3f(r,a^\ast)}{2\eta (a^*) r}
\]
where $f(r,a^\ast)$ is the relativistic (NT) version
of the stress-free inner boundary condition, which depends only on
dimensionless radius $r=R/R_g$ ($R_g=GM/c^2$), and spin, $a^\ast$, and
$f\to 1$ as $r\to \infty$.  The total luminosity integrated over the
entire disc is $L=\eta(a^*)\dot{M}c^2$ where efficiency $\eta(a^\ast)$, and we
use the efficiency factor to define
$\dot{m}=\dot{M}/\dot{M}_{\rm Edd}$ where $\dot{M}_{\rm Edd}=L_{\rm
  Edd}/(\eta c^2)$. Hence $L/L_{\rm Edd}$ is equal to $\dot{m}$ for sub-Eddington discs
of any spin.

We can define a local Eddington flux limit assuming
a spherically symmetric, static geometry of 
$F_{\rm Edd}(R)=L_{\rm Edd}/(4\pi R^2)$. The local disc flux reaches this value for
$F_{\rm NT}(R)= F_{\rm Edd}(R)$, i.e. at 
$$\frac{L}{L_{\rm Edd}}= \dot{m}=\frac{2\eta(a^\ast) r}{3f(r,a^\ast)}$$ 

\begin{figure}
	\includegraphics[angle=-90,width=0.95\columnwidth]{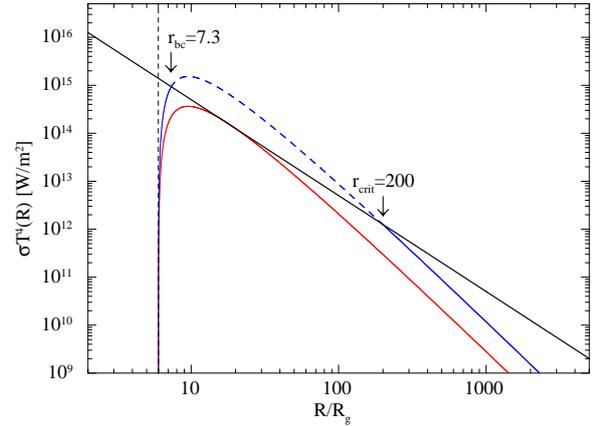}
    \caption{Emissivity is plotted against $r$ fir a $10^7~M_\odot$ black hole. $F_{\rm NT}$ for $a^\ast=0$ of $\dot{m}=2.39$ and 10 are shown with red solid line and blue solid (or dashed) line, respectively. $F_{\rm Edd}$ is shown with black solid line. $r_{\rm isco}$ is indicated with a vertical dashed line.}
    \label{fig:flux}
\end{figure}

Here, we first discuss the case for a non-rotating black hole.
Figure~\ref{fig:flux} shows the NT flux for $\dot{m}_{\rm crit}$ (red),
compared to the local Eddington flux (black). The flux first touches
the local Eddington limit close to the peak of the disc emissivity,
at $r=17.5$ at a critical mass accretion rate of $\dot{m}=\dot{m}_{\rm crit}=2.39$.
At this radius, the curvature is dominated by the stress-free inner boundary
condition which keeps the slope almost parallel to the $R^{-2}$
Eddington flux curve rather than being dominated by the large radius
behaviour of $F_{\rm NT}\propto R^{-3}$.  This simplistic derivation of
the critical mass accretion rate is (surprisingly) the same as
$\dot{m}_{\rm crit}=2\sim 3$ identified by Fig.~3 of \cite{watarai2000} and Fig.~4.11 of \cite{sadowski2011}, on the basis of
more detailed calculations which include the full two dimensional disc
structure including rotation and pressure terms.

\begin{figure}
	\includegraphics[angle=-90,width=0.95\columnwidth]{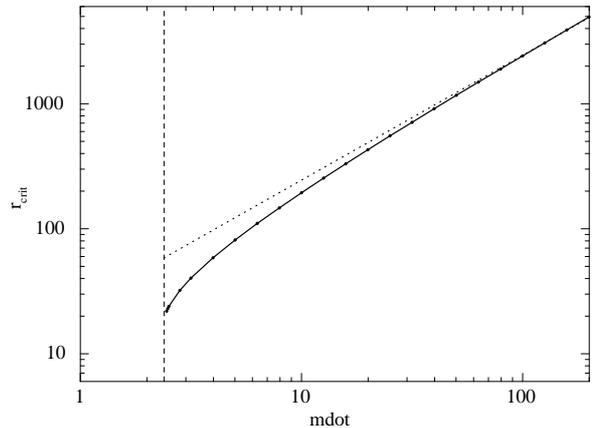}
    \caption{The calculated critical radius $r_{\rm crit}$ is plotted against $\dot{m}$. Solid line indicates the result by $4\pi \sigma R_{\rm crit} ^2  T(R_{\rm crit}) ^4=L_{\rm Edd}$, Dotted line represents $r_{\rm crit}=24.5\dot{m}$. The dashed vertical line indicates $\dot{m}=2.39$.}
    \label{fig:rcri}
\end{figure}

For $\dot{m}>\dot{m}_{\rm crit}$, the local emissivity of the inner
disc becomes less effective than that of NT, as it is limited to the
local Eddington flux. The blue line on Fig.~\ref{fig:flux} shows the result of this simplistic expectation for
$\dot{m}=10$.  The radius at which the flux first reaches the local
Eddington limit is now much larger, at $r\simeq 200$, 
and the inner disc is less luminous than 
predicted from the outer disc assuming NT emissivity.
Figure~\ref{fig:rcri} shows this critical radius, $r_{\rm crit}$ 
at which the flux starts to hit $F_{\rm Edd}$ plotted
against $\dot{m}$ for $a^\ast=0$. 
For $\dot{m}\gg \dot{m}_{\rm crit}$, the critical
radius occurs away from the emissivity peak, so the behaviour
asymptotes to the expected linear relation between $r_{\rm crit}$ and
$\dot{m}$ predicted by the intersection of the intrinsic dissipation
$\propto L/R^3$ and the local Eddington flux $\propto L_{\rm Edd}/R^2$. We
find $r_{\rm crit}=24.5\dot{m}$ for $\dot{m}\gg\dot{m}_{\rm
  crit}$. This differs only by factors of order unity to those of
other calculations e.g. our $r_{\rm crit}$ is $1.5\times $ smaller than
that of \cite{fukue2004}, but is $2\times$  larger than that of
\cite{ss73}. There is a much faster increase in $r_{\rm crit}$ with
$\dot{m}$ close to $\dot{m}_{\rm crit}$ due to the almost parallel slope
of the Eddington flux and emissivity (see Fig. \ref{fig:flux})

Advection also changes the structure of the very inner disc, close to the innermost stable circular 
orbit. Figure~\ref{fig:flux} also shows that the flux at $\dot{m}=10$ drops
below the local Eddington limit for $r< r_{\rm bc}(=7.3)$ due to the stress
free inner boundary condition. At this point we might expect that we
return again to $F_{\rm NT}$ but more detailed models of super-Eddington
discs show that both the inner boundary and the emissivity at small
radii change systematically above $\dot{m}_{\rm crit}$. 
We chose to base our model on $\dot{m}-L$ plots of \cite{watarai2000} and \cite{sadowski2011}.
We find that we obtain consistent $\dot{m}-L$ with their plots for both 
spin 0 and 0.9 
when the inner radius $r_{\rm in}$ decreases from $r_{\rm isco}$ to $r_{\rm h}$ (the horizon) above $\dot{m}=6$ as:
\begin{equation}
\label{eq:rin}
r_{\rm in}=
\begin{cases}
r_{\rm isco}&(\dot{m}\le 6)\\
r_{\rm isco}\cdot  \left(\frac{\dot{m}}{6}\right)^{\log(\frac{r_{\rm h}}{r_{\rm isco}})/\log(\frac{100}{6})}&(6<\dot{m}\le 100)\\
r_{\rm h}&(100<\dot{m})
\end{cases}
\end{equation}
This increases the very inner disc emissivity, so there is more flux
emitted than predicted by $F_{\rm NT}$ in this region where the NT disc
has its emission strongly suppressed by the stress-free inner boundary
condition. We again base our approach to approximately match to
\cite{watarai2000} for spin 0, and \cite{sadowski2011} for spin 0 and 0.9. 
We define the inner radius at which the 
flux drops below the local Eddington flux due to the boundary condition as $r_{\rm bc}$. 
Then for $\dot{m}>\dot{m}_{\rm crit}$ we define the emissivity at $r_{\rm in}<r<r_{\rm bc}$ as:
\begin{equation}
\label{eq:emm}
F_{\rm slim}=
\begin{cases}
F_{\rm NT} \left(\frac{\dot{m}}{\dot{m}_{\rm crit}}\right)^{\log(\frac{F_{\rm Edd}}{F_{\rm NT}})/\log(\frac{6}{\dot{m}_{\rm crit}})}
 & \dot{m}_{\rm crit}< \dot{m}< 6 \\ 
F_{\rm Edd}&\dot{m}\ge 6
\end{cases}
\end{equation}

\begin{figure}
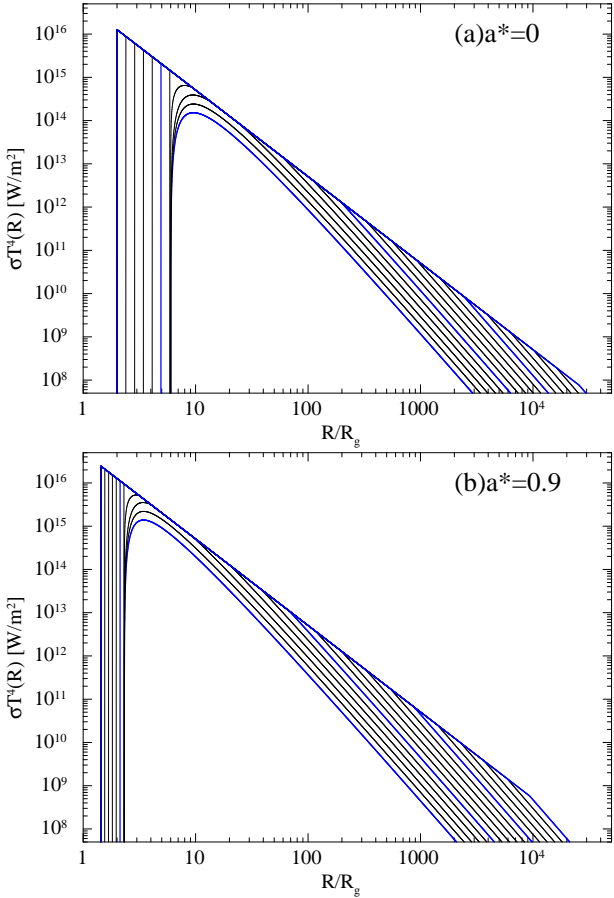

	\includegraphics[angle=-90,width=1\columnwidth]{Figure3a.ps}
	\includegraphics[angle=-90,width=1\columnwidth]{Figure3b.ps}	
    \caption{The local flux, $F(R)=\sigma T_{\rm eff}(R)^4$
    is plotted against radius $r$ for a black hole of 
    $M=10^7M_\odot$ with different $\dot{m}$; 
    from $\log \dot{m}=0$ to $3$ with $\Delta \log\dot{m}=0.2$. Thick blue solid lines represent 
    $\log \dot{m}=0$, 1, $2$ and 
    $3$.  Black hole spin is set to be $a^\ast=0$ (panel a) and 0.9 (panel b).}
    \label{fig:tr}
\end{figure}

This sets the local flux at all radii as a smoothly varying function
with $F_{\rm slim}\le F_{\rm Edd}$, giving a local effective blackbody
temperature of $\sigma T^4_{\rm eff}=F_{\rm slim}$. All these
approximations to the local emissivity are included in
Fig. \ref{fig:tr}a, where we calculate the flux for a $10^7M_\odot$
black hole with $a^*=0$. 
Advection makes the emissivity profile
flatter than that of NT disc, and the emissivity changes smoothly in
the inner region for $\dot{m}=\dot{m}_{\rm crit}=2.39\to 6$, as
$r_{\rm in}$ decreases from $r_{\rm isco}\to r_h$ (i.e. $6\to 2$ for spin 0) for
$\dot{m}=6\to 100$. 
Figure~\ref{fig:tr}b shows the corresponding fluxes for
$a^*=0.9$. There is now much less difference in inner disc radius, as
$r_{\rm isco}=2.3$ and  $r_{\rm h}=1.3$. 

\begin{figure}
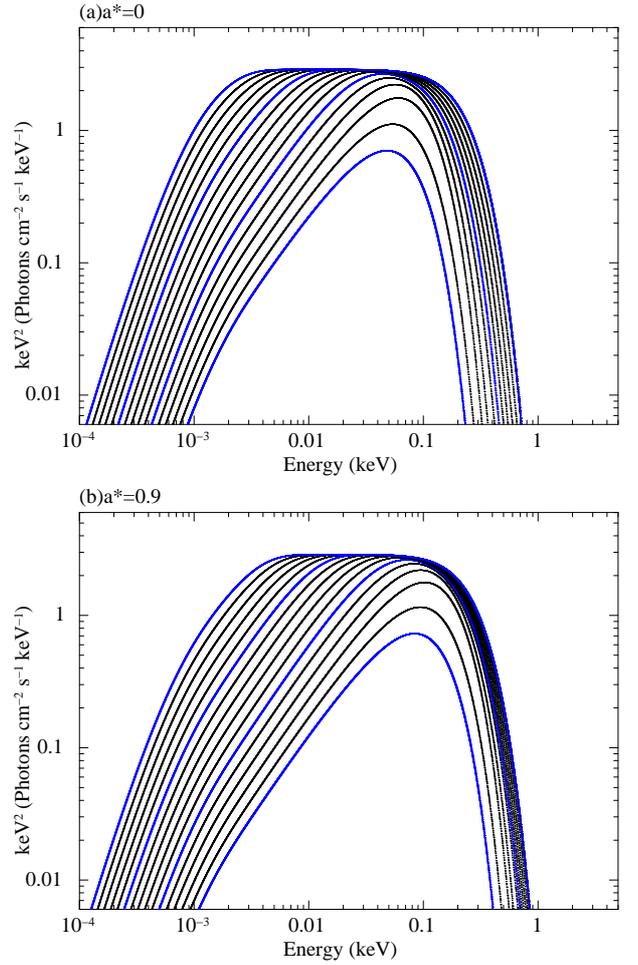

	\includegraphics[angle=-90,width=1\columnwidth]{Figure4a.ps}
	\includegraphics[angle=-90,width=1\columnwidth]{Figure4b.ps}
    \caption{The emergent spectra for a $10^7~M_\odot$ black hole with different $\dot{m}$ from $\log\dot{m}=0$ to $3$ with 
    $\Delta \log \dot{m}=0.2$. Distance and inclination  are set to be 100~Mpc and $0^\circ$, respectively. 
    Colours are the same as Fig.~\ref{fig:tr}}
    \label{fig:slimsed-disc}
\end{figure}

\begin{figure}
	\includegraphics[angle=-90,width=1\columnwidth]{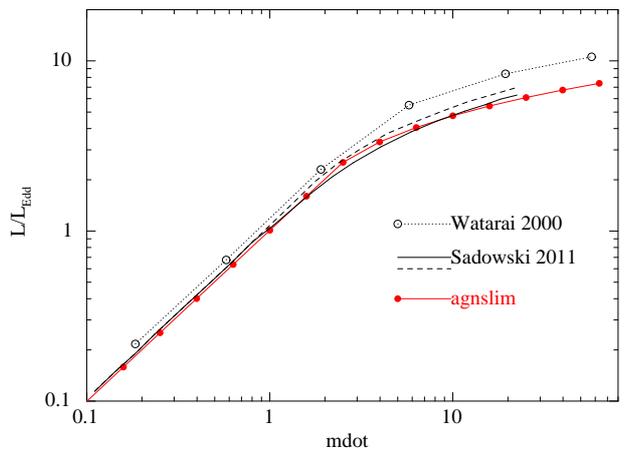}
    \caption{Eddington ratio as a function of $\dot{m}$ based on {\sc agnslim} for spin 0. Data points of {\sc agnslim} and Fig.3 of \citet{watarai2000}  are shown with filled red circles and open black circles, respectively. A solid line and a dashed line is theoretical prediction by \citet{sadowski2011} in Fig.~4.11, with $\alpha=0.01$ and $0.1$.}
    \label{fig:mdot-ledd}
\end{figure}

Figure~\ref{fig:slimsed-disc} shows the resulting spectra, again for
$a^*=0$ (panel a) and $a^*=0.9$ (panel b).  We integrate each spectral
model to derive a bolometric luminosity as a function of $\dot{m}$.
Figure~\ref{fig:mdot-ledd} shows these for $a^*=0$ compared to the more
detailed calculations by  \cite{watarai2000} and \cite{sadowski2011}. Their data
(taken from Fig. 3 in \cite{watarai2000} and Fig.~4.11 in \cite{sadowski2011} 
using {\sc GraphClick}\footnote{http://www.arizona-software.ch/graphclick})
are rescaled from their definition of mass accretion rate onto our $\dot{m}$.

\begin{figure}
	\includegraphics[angle=-90,width=1\columnwidth]{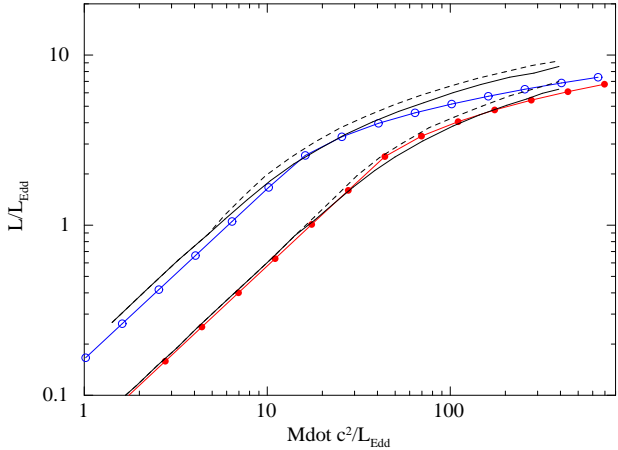}
    \caption{Eddington ratio as a function of $\dot{M}c^2/L_{\rm Edd}$ based on {\sc agnslim}. 
    Same as Fig.~\ref{fig:mdot-ledd}, solid lines and dashed lines are theoretical predictions by \citet{sadowski2011} in Fig.~4.11, with $\alpha=0.01$ and $0.1$. 
   Case of black hole spin of $a^\ast=0$ and 0.9 are shown with red filled circles and blue open circles, respectively. }
    \label{fig:mdot-ledd2}
\end{figure}

It is clear that our model reproduces the overall
bolometric luminosity within the uncertainties of previous
calculations, and in particular is closest to the best current
calculations for spin 0 \citep{watarai2000,sadowski2011}.  
\citet{watarai2000} do not include spin, so we only use
\citet{sadowski2011} to compare with our spin 0.9 results.
Figure~\ref{fig:mdot-ledd2} shows our model $L/L_{\rm Edd}$  
versus $\dot{M}c^2/L_{\rm Edd}$ as used by \cite{sadowski2011}.
Rather surprisingly, our models for $a^*=0.9$ 
show slightly less emissivity at
sub-Eddington rates than \cite{sadowski2011}. 
This is because they do not set $r_{\rm in}=r_{\rm isco}$ for sub-Eddington rates. Instead, they use the 
(viscosity dependent) radius where the 
effective potential forms a self crossing Roche lobe (potential spout). 
For spin of 0.9
this gives $r_{\rm in}=1.96$, smaller than $r_{\rm isco}=2.32$, 
(hence efficiency of  $\sim 0.19$ rather than $0.15$) for 
$\dot{m}>0.3$ \citep{abramowicz2010}. Given that
this radius is dependent on viscosity, and the emission from these inner
radii will be strongly affected by gravitational redshifts and the unknown disc geometry, we choose
to keep $r_{\rm in}=r_{\rm isco}$ below $\dot{m}=6$, as in equation 1.

In summary, advection changes the structure of the inner disc for
$r\le r_{\rm crit}$. It decreases the inner disc emissivity from
$r_{\rm bc}<r<r_{\rm crit}$, but increases it slightly in the very innermost
region, $r_{\rm in}<r<r_{\rm bc}$. The decrease is always more important, so
the inner region of the slim disc is always less luminous than
predicted by NT emissivity. This removes a key spin diagnostic from
energetics. \citet{done2013} show how the observed outer disc
optical/UV emission determines the mass accretion rate, and so
predicts the inner disc luminosity modulo black hole spin.  Higher
spin gives a smaller inner disc radius, so more emissivity, so the
observed inner disc emission can be used to determine black hole spin.
Some extreme AGN have inner discs which are underluminous even for
zero spin assuming NT emissivity, so these strongly constrain spin to
be low \citep{done2016,jin2016,jin2017}. However, these AGN are most probably
super-Eddington, where the inner disc emissivity is strongly reduced, which 
removes the constraint on low black hole spin. 

Advection also removes the other black hole spin diagnostic, namely
the position of the inner disc edge as seen via relativistic smearing
of the iron line.  The disc inner edge is now set by pressure forces
as well as gravity, and can be inwards of the innermost stable
circular orbit. Thus a very broad iron line, implying a very small
inner radius, does not necessarily imply high spin for a super-Eddington flow. 

\section{Spectral Model}

\begin{figure}
	\includegraphics[width=0.95\columnwidth]{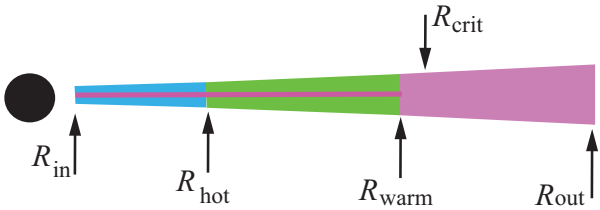}
    \caption{Geometry}
    \label{fig:geometry}
\end{figure}

The sub-Eddington accretion flow model {\sc agnsed} (KD18) 
is able to reproduce the broadband SED of most AGN and quasars
assuming the NT emissivity for a given $\dot{M}$ throughout the disc.
Unlike 'pure' disc models, which assume that the dissipated power
is emitted as blackbody at all radii, it splits the disc into
three different emission regions as schematically shown in
Fig.~\ref{fig:geometry}. The luminosity in the inner region, from
$R_{\rm in}$ and $R_{\rm hot}$, is dissipated in hot 
material, forming a Comptonised spectrum extending up
to high energies, while that from $R_{\rm hot}$ to $R_{\rm warm}$,
is emitted instead in an optically thick, warm Comptonising region,
and only the outer regions, from
$R_{\rm warm}$ to $R_{\rm out}$, completely thermalise and
emit as a standard blackbody.

We follow this general model, but replace the NT emissivity with the
slim disc emissivity detailed above. 
Though the slim disc is expected to give larger $H/R$, it is complicated to include this geometrical effects to the model.
We thus still employ the thin disc geometry as is done for 
{\sc agnsed}. 
One key difference is that the
{\sc agnsed} model assumes that the hot inner flow region does not
have an underlying disc component. This is because the lowest
luminosity AGN have hard X-ray spectra with photon index $\Gamma<1.9$.
Reprocessing of the coronal X-rays in any underlying cool disc
material provides seed photons for the hot Comptonisation and sets
$\Gamma=1.9$ as a lower limit for the spectral index even with a
non-emitting 'passive' disc. While the higher luminosity AGN typically
have X-ray spectra which are softer than this limit, the observed
spectral indices are remarkably well predicted for $0.03<L/L_{\rm Edd}<1$
assuming this truncated disc geometry. However, for super-Eddington
flows it seems very unlikely that this continues (see Section 2) so we
consider a slab corona for the hard X-rays at these high mass
accretion rates instead truncated disc with spherical hot flow.

Another, more minor difference from {\sc agnsed} is that we do not
include the contribution of reprocessed hard X-ray flux, as we expect
that the inner disc will puff up to some extent (see section 5), so
that there is self-occultation preventing illumination of the outer
disc. Even without this effect, the contribution from reprocessing is
small for high Eddington ratio AGN as they typical have
$L_{\rm X}/L_{\rm bol}\ll 1$ (see Fig.3 in KD18).

\begin{figure}
	\includegraphics[angle=-90,width=1\columnwidth]{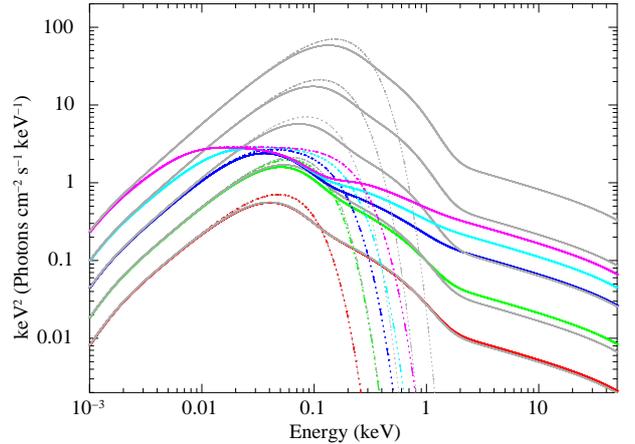}
    \caption{Comparison of {\sc agnsed} (gray) and {\sc agnslim} (colour) for $M=10^7~M_\odot$ black hole with $i=0^\circ$. $\dot{m}=1$ (red), 3 (green), 10 (blue), 30 (cyan) and 100 (magenta) are shown
     with $kT_{e,{\rm warm}}=0.2$~keV, $kT_{e,{\rm hot}}=100$~keV, $\Gamma_{\rm warm}=2.7$, $\Gamma_{\rm hot}=2.4$,
    $r_{\rm hot}=10$ and $r_{\rm warm}=20$. Intrinsic disc emissions without any Comptonisation are shown with dotted lines.}
    \label{fig:slimsed-sc}
\end{figure}

We name this new model {\sc agnslim}.
We show a comparison of {\sc agnslim} (coloured lines)
versus {\sc agnsed} (grey) in Fig.~\ref{fig:slimsed-sc} for
$\dot{m}=1$, 3, 10, 30 and 100. The dot-dashed lines show the pure
intrinsic blackbody emission without any Comptonisation, while the solid lines show the resulting spectra
for a set of fiducial parameters which are characteristic of high
$\dot{m}$ AGN ($\Gamma_{\rm warm}=2.7$, $kT_{e, \rm{warm}}=0.2$~keV, 
$r_{\rm warm}=20$, $\Gamma_{\rm hot}=2.4$, $kT_{e, \rm{hot}}=100$~keV and
$r_{\rm hot}=10$ for a $10^7M_\odot$ black hole with spin 0 viewed at
$0^\circ$ inclination).  There is no visible difference between {\sc
  agnsed} and {\sc agnslim} at $\dot{m}=1$ (red) as this is below
$\dot{m}_{\rm crit}=2.39$, so the local flux is set by the standard NT
emissivity and never hits the saturation point at the local Eddington
limit.  The different geometry for the hot X-ray regions (disc corona
rather than truncated disc) does make a difference to the high energy
X-ray flux, but the two are equal for the face on inclination assumed
here.  There is a slight difference at $\dot{m}=3$ (green) as this
just touches the local Eddington limit, so has flux saturated at this
limit from $r=35\to7$. This produces a noticeable drop in the hottest
blackbody disc emission ($r=35\to 20$) and in the warm Comptonisation
emission from $r=20\to 10$. However, the hot Comptonisation in {\sc
  agnslim} is slightly brighter than that in {\sc agnsed} due to the
enhanced emissivity from $r_{\rm in}\to r_{\rm bc}$.  The hottest part of the
blackbody disc continues to decrease relative to {\sc agnsed} as
$\dot{m}$ increases from $10$ (blue) to $30$ (cyan) and finally to
$100$ (magenta). This saturation starts to become evident in the UV
for $\dot{m}>10$, and even extends into the optical for $\dot{m}>100$
for this black hole mass. Similarly, the warm Comptonisation emission
continues to decrease relative to {\sc agnsed} to an almost constant
flux in {\sc agnslim} as the region from $r=10\to 20$ is all saturated
to the local Eddington flux. However the hot Compton component shows
more complex behaviour as the increase in emissivity from $r_{\rm bc}\to
r_{\rm in}$ relative to NT can offset some of the decrease from $r=10\to
r_{\rm bc}$.

\label{sec:geometry}

\section{Application to the observed SED of RX~J0439.6-5311}
\label{sec:application}

RX~J$0439.6-5311$ is a type I Narrow Line  quasar
which has the smallest H$\beta$ full-width-at-half-maximum (FWHM) of
$700\pm140~{\rm km~s^{-1}}$ among the 110 soft X-ray selected AGN in
\cite{grupe2004a} and \cite{grupe2004b}. This, together with the
monochromatic luminosity at 5100~\AA~can be used to estimate mass of
$3.9\times 10^6~M_\odot$ (using \citealt{kaspi2000}).  \citet{jin2017}
re-analysed the components in the H$\beta$ line profile and estimate
the broad line width of $\sim 850$~km/s, giving slightly larger mass
estimates of $9.4\times10^6~M_\odot$ \citep{vp06} or $6.7\times
10^6~M_\odot$ \citep{wu02}.

This NLS1 has one of the best broadband datasets of any super-Eddington
AGN, with extremely low galactic absorption column of $N_{\rm
  H}=7.45\times 10^{19}~{\rm cm^{-2}}$, corresponding to reddening
$E(B-V)=1.7\times 10^{-22}N_{\rm H}=0.127$. This means that the AGN
continuum can be seen up to 912 \AA ~in the rest frame, and above
0.1~keV, with less than an order of magnitude gap in energy coverage
from interstellar absorption. We use the optical/UV/X-ray data from
\cite{jin2017}, but exclude the IR as this is dominated by reprocessed
emission from hot dust rather than by the accretion flow itself.

\cite{jin2017} fit these data using the {\sc optxagnf} model, with
redshift fixed at $z=0.242$, i.e. comoving radial distance of
985.0~Mpc and inclination of $i=30^\circ$, together with Galactic
extinction and reddening ({\sc tbnew}, with abundances of
\cite{wilms2000} and {\sc redden}) fixed to the parameters above.
They found that the SED could not be well fit with the NT based
emissivity for black hole masses below $\sim 2\times 10^7M_\odot$. The
lower black hole masses from the virial estimates give SED models which
strongly over predict the observed soft and hard X-ray luminosity for the mass
accretion rates required to fit the optical/UV outer disc continuum. 

Here we first explore the optical/UV continuum alone.
Figure~\ref{fig:slimsed-sc} shows that for extreme $\dot{m}$ the
flux saturation at the local Eddington limit starts to affect even the
optical/UV emission. The advected flux means that there is less flux
emitted so the UV continuum slope becomes significantly redder.  This
should give a secure upper limit of $\dot{m}$ from the optical/UV data
alone, and hence a secure lower limit to the black hole mass. 

\subsection{A secure lower limit for mass from optical/UV data alone}
\label{sec:opt-uv}

As shown in Fig.4 of \cite{jin2017}, the slope of optical-UV data
points in the SED of RX~J$0439.6-5311$ are in good agreement with the
standard NT disc.  There is no strong deviation (flattening) from the
NT disc in the optical/UV range which would be expected if the disc
had become super-Eddington at radii which emit in this bandpass
(Fig.~\ref{fig:slimsed-disc} and \ref{fig:slimsed-sc}).  Thus lack of
such flattening gives us a secure lower limit for $M$ from this upper
limit for $\dot{m}$. We use a pure disc model, as it is only the
X-rays which require the Comptonised components.

We fix the redshift, distance and galactic absorption column at the same values as \cite{jin2017}, but fix 
$i=0^\circ$.
We then fit all the optical-UV data without any Comptonisations with the same parameters except
normalizations of the (non-simultaneous) HST and ground based optical
spectra are multiplied by the best fit factor derived from the broad
band SED spectrum (see next section).  
For a black hole of $a^*=0$, we find an upper limit on
$\dot{m}\simeq 45$, and corresponding lower limit on the mass of
$2.8\times 10^6 M_\odot$. Since we assumed a face on geometry, this can be taken as the secure lower limit for the mass of
RX~J$0439.6-5311$.
We have same lower limit mass of $2.8\times 10^6 M_\odot$ with higher 
$\dot{m}$ of $\sim120$ for a black hole of $a^*=0.9$. This is because the saturated flux does not depend on black hole spin but on black hole mass.

\begin{figure}
\includegraphics[angle=-90,width=1\columnwidth]{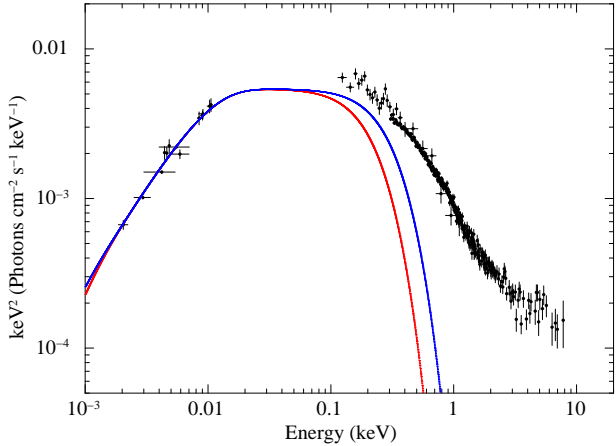}
    \caption{Absorption corrected broadband SED of RX~J$0439.6-5311$ and {\sc agnslim} of lower limit mass $M=2.8\times 10^6~M_\odot$ without warm/hot Comptonising corona. The SED is deconvolved based on the best fit model in section~\ref{sec:broadband}. Blackhole spin is assumed to be $a^\ast=0$ (red) and 0.9 (blue).}
    \label{fig:eeu_opt}
\end{figure}

Red line and blue line in Fig.~\ref{fig:eeu_opt} shows the pure disc {\sc agnslim} model for a black hole of
$M=2.8\times 10^6~M_\odot$ with $\dot{m}=45$ ($a^*=0$) and $\dot{m}=120$
($a^*=0.9$), respectively. The absorption corrected
broad band data are overlaid on this figure, clearly showing that the
higher energy spectra cannot be explained by a slim disc model where
the emission is blackbody. It is also clear that this minimum mass
model for the optical/UV does not have enough luminosity to power the
observed emission, so a higher black hole mass/lower mass accretion
rate with lower advective losses is required in order to match the
data.

\subsection{Broadband SED fit }
\label{sec:broadband}

We now fit the entire broadband SED with {\sc agnslim} model.  We again
renormalize the optical-UV data to that of XMM-OM data. We first allow the
normalisation of the {\it ROSAT} data to be free relative to the 
XMM EPIC-pn data over the same bandpass of 0.3--1keV, but the 
resulting ratio was $1.02^{+0.07}_{-0.08}$, so we
fixed this at 1.

\begin{figure}
\includegraphics[angle=-90,width=1\columnwidth]{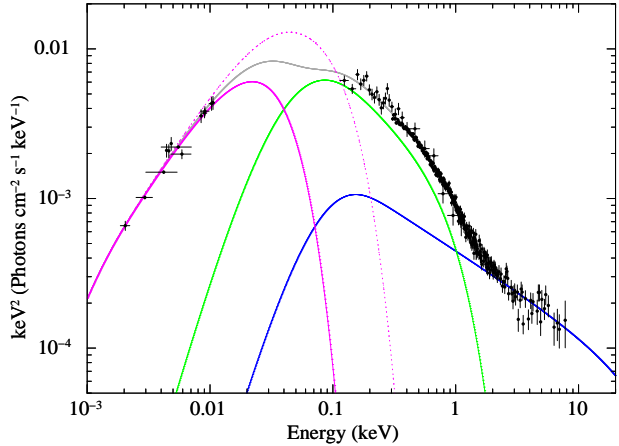}
    \caption{Same as Fig.~\ref{fig:eeu_opt} but with the best fit {\sc agnslim} model without GR for a black hole mass 
$ 8\times 10^6~M_\odot$ and $a^\ast=0$ (grey), showing the
outer blackbody disc (magenta), warm comptonised disc-corona (green)
and hot inner disc-corona (blue). The dashed magenta line shows the 
same slim disc model but with pure blackbody emission 
at all radii.}

    \label{fig:eeu_rxj04}
\end{figure}

The {\sc agnslim} model with $a^\ast=0$ reproduces the broadband SED fairly well, with
the best fit $\chi^2_\nu$ around 1.30 for a range of black hole
masses, from $5 \times 10^6~M_\odot$ to $1.4\times 10^7~M_\odot$,
corresponding to $\dot{m}\simeq 14$ to $2$. The high mass/low mass accretion rate fit is 
consistent with the fits of \cite{jin2017}, as $\dot{m}=2<\dot{m}_{\rm crit}$ so does not 
reach the local Eddington flux, so still has NT emissivity as assumed in the models used in \cite{jin2017}. However, the {\sc agnslim} model now allows lower mass solutions. 
We fix the mass at $8\times 10^6~M_\odot$ as derived from the H$\beta$ line profile, and  show the best fit parameters in table~\ref{tab:rxj04}. This
model is overlaid on broadband SED in Fig.~\ref{fig:eeu_rxj04}, with the inner hot Comptonising corona (blue), 
the warm Comptonising corona (green) and outer disc (magenta) shown separately.
The resultant accretion rate of $\dot{m}=5.4$ is in the slim disc regime, and gives an 
observable luminosity of $\sim 2.5L_{\rm Edd}$ (see Fig.~\ref{fig:mdot-ledd}).  
The inner hot corona region is from $r_{\rm hot}=9.4$ to $r_{\rm in}=6$, as $\dot{m}=5.4$ is not high enough to extend the inner disc below $r_{\rm isco}$ (see equation 1). This gives $L_{\rm hot}=0.4L_{\rm Edd}$, which is fairly large compared to some sub-Eddington AGN (KD18).
We repeat the fit for spin of $a^*=0.9$. As shown in table~\ref{tab:rxj04}, the model can fit the data with similar $\chi^2$ to $a^\ast=0$.  Again $r_{\rm in}=r_{\rm isco}$ but the high spin means that
inner part of the flow now extends considerably closer to the central black hole. Thus the
hot inner corona region is more compact, at $r_{\rm hot}=2.7$ to keep the observed $L_{\rm hot}$ of $\sim 0.4L_{\rm Edd}$.
As a result, the warm compact region is also smaller, with $r_{\rm warm}=6.1$.

However, in both these fits the soft and hard X-ray emission regions are close enough to the black hole that
General Relativistic effects should be important \citep{cunningham1975,zhang1997}.
These will be dominated by gravitational and transverse redshift as the face on inclination means that the doppler shift from the orbital velocity is small. The redshifts reduce the observed luminosity, but in a way which 
is difficult to calculate exactly as 
slim discs are no longer flat or Keplarian,
so the ray tracing depends on the poorly known geometry and dynamics of the flow  \citep{vierdayanti2013}.
Even for sub-Eddington, flat discs, these
effects are not easy to include as the GR transfer functions available in {\sc xspec} are normalised to unity, rather than incorporating the radially dependent 
loss of photons down the black hole. Instead, we follow the approximate treatment suggested in \cite{done2013}, using the standard {\sc xspec} transfer function {\sc kdblur} from $r_{\rm in}$ to $r_{\rm hot}$ on the hot component, 
 $r_{\rm hot}$ to $2r_{\rm hot}$ on the warm comptonisation, and $r_{\rm warm}$ to $2r_{\rm warm}$ on the outer disc.
 This at least gives some indication of the size of the
uncertainties introduced by GR effects (see also \citealt{porquet2019}). We show results for  $a^*=0$ and $0.9$  in table~\ref{tab:rxj04}. The best fit for $a^*=0$ and black hole of  mass $8\times 10^6~M_\odot$ is now considerably worse than before, at  $\chi^2/dof=948/711$. This is because the GR effects reduce both the soft and hard X-ray flux. The mass and spin are fixed in this fit, and $\dot{m}$ is tightly constrained by the optical/UV data, and even increasing $r_{\rm hot}$ and $r_{\rm warm}$ does not give enough X-ray flux. Allowing the black hole mass to be free does allow us to recover a fit including GR  of comparable quality to one without, but requires a mass of
$>10^7~M_\odot$, and hence $\dot{m}<\dot{m}_{\rm crit}$ out of the slim disc regime. There are similar 
issues with including GR at higher spin, but the stronger suppression of the soft and hard X-ray flux
is less of a problem as $r_{\rm hot}$ and $r_{\rm warm}$ were so small without GR that they can increase
enough to compensate for the redshift losses. 

%


\begin{table*}
	\centering
	\caption{The best fit parameters of RXJ~$0439.6-5311$. }
	\label{tab:rxj04}
	\begin{tabular}{lccccc} 
		\hline
&& $a^\ast=0$ without GR&$a^\ast=0.9$ without GR& $a^\ast=0$ with GR$^\dagger$ &$a^\ast=0.9$ with GR$^\dagger$\\
\hline
$\dot{m}$ &&$5.43^{+0.12}_{-0.11}$&$13.7\pm0.3$&$5.61^{+0.14}_{-0.11}$&$13.1^{+0.5}_{-0.2}$\\
$\Gamma_{\rm hot}$&&$2.53\pm0.07$&$2.55\pm0.07$&$2.60^{+0.04}_{-0.06}$&$2.49^{+0.05}_{-0.07}$\\
$\Gamma_{\rm warm}$ &&$2.69^{+0.04}_{-0.05}$&$2.67\pm0.07$&$2.58\pm0.03$&$2.67^{+0.04}_{-0.02}$\\
$kT_{e,{\rm warm}}$ &keV&$0.252^{+0.013}_{-0.012}$&$0.247^{+0.016}_{-0.014}$&$0.253^{+0.011}_{-0.006}$&$0.330^{+0.021}_{-0.013}$\\
$r_{\rm hot}$& &$9.4^{+0.7}_{-0.5}$&$2.71^{+0.09}_{-0.19}$&$12.0^{+1.0}_{-0.9}$&$4.2\pm0.3$\\
$r_{\rm warm}$& &$60^{+9}_{-7}$&$6.11^{+0.11}_{-0.21}$&$100^{+3}_{-1}$&$53^{+14}_{-8}$\\
\hline
\multicolumn{6}{l}{The other size scales calculated  based on the best fit model}\\
\hline
$r_{\rm in}$& &6.0&2.0&6.0&2.0\\
$r_{\rm crit}$& &91&100&94&95\\
$r_{\rm out}$& $\times 10^3$&$8.0$&$12.1$&$8.1$&$11.9$\\
\hline
$\chi^2/{\rm dof}$&&922.2/711&924.6/711&948.0/711&926.9/711\\
\hline
\multicolumn{6}{l}{Electron temperature of the hot comptonising corona is fixed at 20~keV.}\\
\multicolumn{6}{l}{Black hole mass is fixed at $8.0\times 10^6~M_\odot$.}\\
 \multicolumn{6}{l}{$^\dagger$ General relativistic effects are approximated by {\sc kdblur} with power index of 3. The model is described as}\\
 \multicolumn{6}{l}{ {\sc kdblur(h)*agnslim(hot compton)+kdblur(w)*agnslim(warm compton)+kdblur(o)*agnslim(outer disc)}.}\\
  \multicolumn{6}{l}{$r_{\rm in}$ of {\sc kdblur(h)} is fixed at 2 and 6 for $a^\ast=0$ and 0.9, respectively, and $r_{\rm out}$ of {\sc kdbur(h)} is fixed at $r_{\rm hot}$.}\\
  \multicolumn{6}{l}{$r_{\rm in}$ and $r_{\rm out}$ of {\sc kdblur(w)} are fixed at $r_{\rm hot}$ and $2r_{\rm hot}$, respectively.}\\
  \multicolumn{6}{l}{$r_{\rm in}$ and $r_{\rm out}$ of {\sc kdblur(o)} are fixed at $r_{\rm warm}$ and $2r_{\rm warm}$, respectively.}\\
	\end{tabular}
\end{table*}

\section{Application to $z=6.621$ quasar PSO~J$006+39$}

High mass accretion rates are expected in the gas rich early Universe,
but so far there is only one AGN known at $z>6$ which is strongly
super-Eddington. This is the recently discovered quasar PSO~J$006+39$,
at $z=6.621$ \citep{tang2017}, with a black hole mass measured from
the width of the Mg$_{II}$ line of $(1.4$--$1.7)\times 10^8~M_\odot$
\citep{tang2019}.  This paper also showed that the rest frame
optical/UV continuum was extremely blue, with $\alpha_\nu=0.94$ (i.e,
photon index $\Gamma=0.06$), significantly bluer than the slope of
the standard disc of $\alpha_\nu=1/3$ (i.e., $\Gamma=2/3$), and
significantly bluer than the majority of quasars at either high $z$
(\citealt{tang2019} and references therein) or more locally
\citep{xie2016}. A slope which is redder than the standard disc is
easy to obtain (dust reddening in the host galaxy: \citealt{baron2016}, Davis et al. 2007
; or a colour temperature correction onset in the UV: \citealt{done2012},
or advection losses, see sections 3 and 4). However, a bluer slope is
more difficult to explain. \cite{tang2019} showed that this could be
obtained if the disc is significantly smaller than expected from the
self gravity radius of $r_{\rm out}\sim 5000$. They fit PSO~J$006+39$
using a model where the outer edge of the accretion disc is at $r_{\rm
  out}~\sim230$, with a highly super-Eddington rate of $\dot{m}\sim9$. 

These fits were derived using the {\sc optxagnf} model in {\sc xspec},
which is similar to {\sc agnsed} in assuming NT emissivity. Yet this is not 
self consistent as at such
high $\dot{m}$ optically thick advection must be important.  We thus
fit the optical/UV continum spectrum with our {\sc agnslim} model.  We
use the data from \cite{tang2019} which is corrected for redshift and
for Galactic absorption, and fix the black hole mass 
 at $1.6\times 10^8~M_\odot$. Since the data have been redshift
corrected we use the luminosity distance fixed at 66.1~Gpc to derive
flux. We assume a pure blackbody disc model since there are (currently)
no published X-ray data from which to determine the hot coronal component. 
We neglect GR as the (rest frame) optical/UV data are not from the inner disc where 
these effects become important. 

\subsection{{\sc agnslim} model fits}

\begin{figure}
\includegraphics[angle=-90,width=1\columnwidth]{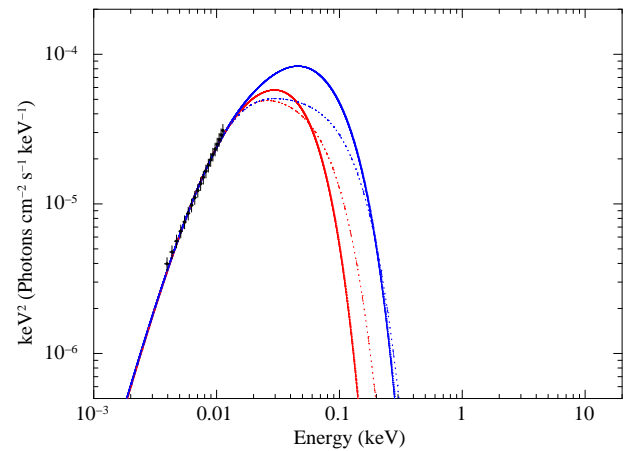}

    \caption{The best fit {\sc agnslim} model overlaid on the redshift corrected NIR spectrum of PSO~J$006+39$. Solid lines and dotted lines correspond to $i=0^\circ$ and $60^\circ$, respectively. Spin parameter $a^\ast$ is assumed to be 0 (red) and 0.9 (blue) . 
        } 
    \label{fig:eeu_p006}
\end{figure}

We first fit the data by setting $r_{\rm out}$ to the value determined
by the self-gravity and we fix $\cos i=1$. We confirm that this gives a very poor fit, with
$\chi^2/dof=171/19$ and 123/19 for $a^\ast=0$ and 0.9, respectively,
showing that the data are much bluer than the model, as before.  We
then follow \cite{tang2019} and allow $r_{\rm out}$ to be a free
parameter. 
The results are shown in table~\ref{tab:pso06} and Fig.~\ref{fig:eeu_p006}.
The spectrum is well reproduced
by the model with $\dot{m}=2.2$ and $r_{\rm out}=140$ for spin 0, and for 
$\dot{m}=4.4$ and $r_{\rm out}=145$ for spin 0.9.
These values of $r_{\rm out}$ are 
consistent with the result of \cite{tang2019} as the
difference between NT based emissivities and the slim disc emissivity
is not large in this regime. Thus our fits confirm the requirement for
an extremely small disc outer radius in this super-Eddington object.

\begin{figure}
	\includegraphics[width=0.95\columnwidth]{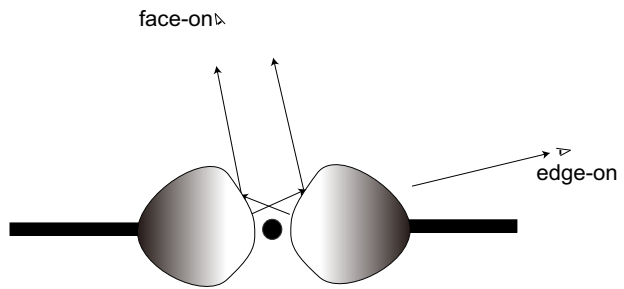}
    \caption{Schematic picture of puffed slim disc.}
    \label{fig:slim_puff}
\end{figure}

Geometrical effects may give a solution for this difference in outer
radius, as proposed by \cite{tang2019}. The inner slim disc region
should be puffed up in highly accreting black holes (see
Fig.~\ref{fig:slim_puff}). Viewing angles which intersect the opening
angle of the inner disc will have the emission from $r_{\rm in}\to
r_{\rm crit}$ geometrically boosted by a factor $\sim
(1-\sin\theta_H)^{-1}$ where $\tan\theta_H=H/R$. Thus it may appear
that the disc itself ends at $r_{\rm crit}$ as this geometric beaming
makes the inner disc emission brighter than that of the outer disc
beyond $r_{\rm crit}$. This is an initially very attractive solution as
the values of $r_{\rm out}$ determined above are very similar to the
values of $r_{\rm crit}$ expected for these mass accretion rates.

However, any geometric beaming reduces the required mass accretion
rate.  The radii emitting the optical/UV flux is still mostly in the
standard NT regime so gives flux $\propto (M\dot{M})^{2/3}\propto
(M^2\dot{m})^{2/3}$. Hence enhancing the flux within the funnel by a
factor $b$ will decrease $\dot{m}$ by a factor $b^{3/2}$. 
Thus, large beaming factors mean that the underlying flow
here would not be super-Eddington, in conflict with the assumption that
the super-Eddington flux has caused the disc to puff up. 
However slim discs probably do not form extreme funnels, they have $H\lesssim R$
for $\dot{m}\gg 1$ \citep{abramowicz88,lasota2016}. 
Thus these should be only mildly beamed  by less than a factor of 3-4 (see also \citealt{atapin2019}), although any wind may enhance the
funnel geometry beyond this (e.g., \citealt{abolmasov2009, poutanen2007}). The unbeamed {\sc agnslim} fits required $\dot{m}=2\to 4$ for $a^*=0\to 0.9$ so, 
{\sc agnslim} even with mild beaming with $b=3$  (i.e. fixing the normalisation  to this value)
means that $\dot{m}$ is reduced to 
below Eddington, in conflict with the model where the disc puffs up
due to a strongly super-Eddington mass accretion rate.
Thus our  conclusion is that it is difficult to explain the observed very blue
continuum in PSO~J$006+39$ by emission dominated by the inner funnel
of a super-Eddington flow. Stronger geometric beaming will reduce the
intrinsic $\dot{m}$ still further, making an even larger discrepancy
between $r_{\rm out}$ as required from fitting the data, and $r_{\rm crit}$
inferred from $\dot{m}$.

The failure of the previous models to give a fully self consistent
picture motivates us to consider a more subtle effect of the geometry
shown in Fig.~\ref{fig:slim_puff}. The far side of the puffed up disc
will be enhanced relative to the outer standard disc as it makes a
smaller angle to the line of sight (Kawashima T. in private communication; see also \citealt{wang2014} for
similar geometric effects on the illumination of the BLR by a
super-Eddington flow). We fix the inclination to $60^\circ$ to approximate  spherical geometry for far side of the puffed up disc,
and fit the data for both $a^*=0$ and $0.9$. The results are again shown in table~\ref{tab:pso06} and Fig.~\ref{fig:eeu_p006}.
These again give
small outer disc radii of $\sim 190R_g$,  This value is almost consistent with 
$R_{\rm cri}$ for $\dot{m}\sim 10$ ($a^*=0$) or for $\dot{m}\sim20$ ($a^*=0.9)$.  Yet here
to enhance the edge of puffed flow than the redder outer standard disc by this geometrical effect, 
extremely high inclination angle of $i=80^\circ\sim90^\circ$ is required. Thus 
it seems unlikely that the far side of
the puffed up flow would be able to outshine the redder outer disc
emission, so the blue continuum is difficult to explain. Also  it
seems much more likely that any object we see at such high redshift is
the brightest in its class. 

We conclude that the most likely solution for the extremely blue
continuum in PSO J$006+39$ is that the outer disc is intrinsically
small (see also \citealt{collinson2017} for inferred small size outer discs
in quasars at $z\sim 1-2$). Alternatively, we could be looking not at a disc at all, 
but at a wind photosphere which gives approximately blackbody emission.

\begin{table*}
	\centering
	\caption{The best fit parameters of PSO~J$006+39$. }
	\label{tab:pso06}
	\begin{tabular}{lccccc} 
		\hline
&& $a^\ast=0$ ($i=0^\circ$)&$a^\ast=0.9$ ($i=0^\circ$)& $a^\ast=0$ ($i=60^\circ$) &$a^\ast=0.9$ ($i=60^\circ$)\\
\hline
$\dot{m}$ &&$2.2^{+0.6}_{-0.3}$&$4.4^{+1.2}_{-0.7}$&$12^{+174}_{-7}$&$19^{+>1000}_{-7}$\\
$r_{\rm out}$~& &$140^{+21}_{-19}$&$145^{+24}_{-21}$&$189^{+34}_{-23}$&$194^{+34}_{-27}$\\
\hline
\multicolumn{6}{l}{The other size scales calculated based on the best fit model}\\
\hline
$r_{\rm in}$& &6&2.3&4.6&1.90\\
$r_{\rm crit}$& &---&24&226&148\\
\hline
$\chi^2/{\rm dof}$&&3.90/18&3.45/18&4.46/18&4.38/18\\
\hline
\multicolumn{6}{l}{Black hole mass is fixed at $1.6\times 10^8~M_\odot$.}\\
	\end{tabular}
\end{table*}

\label{sec:geometry}

\section{Summary and Conclusions}

Super-Eddington accretion flows are seen in the local Universe, in some
extreme NLS1, ULXs and black hole binary systems, and should be prevalent in
the gas rich conditions of the early Universe. Such flows should be
less radiatively efficient than standard sub-Eddington discs, as the
emitted flux should saturate at the local Eddington limit, with the
remaining power lost through radial advection and/or winds. We develop
a new model for fitting data from super-Eddington flows, {\sc agnslim}.
We base this on the successful sub-Eddington flow model {\sc agnsed}
(an updated version of {\sc optxagnf}) which uses the NT
emissivity for a constant mass accretion rate, but allows radial
stratification of the emission. The outer disc thermalises to a
blackbody, and the inner disc dissipates the emission in a hot,
optically thin Comptonising region, while intermediate radii are
dominated by a warm, optically thick Comptonisation corona.  Our new
model uses this same radial stratification for the spectrum, but
limits the emissivity to the local Eddington flux in order to include
the effects of advection and/or winds.

We show that this model can fit the broadband spectrum of RX
J0439.6-5311, an extreme NLS1 which could not be fit with a
NT emissivity for the most likely black hole masses of
$\sim 8\times 10^6M_\odot$ as its inner disc is clearly much less
luminous than predicted by the high mass accretion rate required to
fit the outer disc emission \citep{jin2017}. Our new model can fit
these data with this low black hole mass, implying $\dot{m}=5\sim 10$.
These slim disc models can no longer constrain black hole spin from
the energetics as both low and high spin models have similar
emissivity and inner disc radii due to the flux saturation at the
local Eddington limit.

Super-Eddington mass accretion rates can solve the problem of the
seed black hole mass required for the highest redshift
quasars (e.g. \citealt{volonteri2005}). 
However, most known quasars at $z>6$ are not
super-Eddington. This could be due to evolutionary effects, where they
are mostly obscured when accreting at such high rates, so that they
only become visible along most sight lines when $\dot{m}\lesssim 1$
(e.g. the review \citealt{alexander2012}).  Nonetheless, there is
one strongly super-Eddington quasar at $z>6.5$, PSO J006+39, which also
has an unusually blue rest frame optical/UV continuum
\citep{tang2019}. We fit this with our new model and show that this
cannot be easily explained even including geometrical effects which
might be expected in the super-Eddington regime. We conclude that these
data require that the disc is intrinsically small rather than being
puffed up, or that the emission is dominated by a wind photosphere.

The model should also be applicable to the accretion discs in ULXs. 
Several of these are now known to be pulsars (ULX-P), with neutron star massess strongly requiring highly super-Eddington 
flows, and the remainder are likely to be black holes with moderately super-Eddington flows \citep{kaaret2017, atapin2019}. 
This model will be publically released as part of the {\sc xspec}
spectral fitting software.

\section*{Acknowledgements}

We thank to C. Jin and J.-J. Tang for providing us a data set for
RX~J$0439.6-5311$ in \cite{jin2017} and PSO~J$006+39$ in
\cite{tang2019}, respectively.  AK acknowledges helpful discussion
with S. Mineshige, J. Fukue and T. Kawashima.  AK is supported in part
from research program in foreign country by Shibaura Institute of
Technology. CD acknowledges support from STFC through grant ST/P000541/1.
We also thank M.~D. Caballero-Garc{\'\i}a as our referee.









\appendix


\bsp	
\label{lastpage}
\end{document}